\documentclass{emulateapj}

\newcommand{\metal}{[Fe/{}H]}
\newcommand{\cfe}{[C/{}Fe]}
\newcommand{\abund}[2]{[#1/{}#2]}
\newcommand{\hes}{{HES}}
\newcommand{\hk}{{HK}}
\newcommand{\ch}{{CH}}
\newcommand{\cn}{{CN}}

\newcommand{\cemp}{{CEMP}}
\newcommand{\ctwo}{{C$_2$}}

\newcommand{\kp}{{KPHES}}
\newcommand{\gp}{{GPHES}}

\newcommand{\jk}{({J$-$K})$_0$}

\newcommand{\gpe}{{GPE}}

\newcommand{\twomass}{{2MASS}}

\newcommand{\kphes}{{KPHES}}
\newcommand{\gphes}{{GPHES}}
\newcommand{\nc}{\newcommand}
\nc{\teff}{$T_{\rm eff}$\,}  
\nc{\logg}{log\,$g$\,}

\shorttitle{A Search for Unrecognized \cemp{} Stars}
\shortauthors{Placco et al.}

\begin{document}

\title{A Search for Unrecognized Carbon-Enhanced Metal-Poor Stars in the Galaxy}

\author{Vinicius M. Placco}
\affil{Departamento de Astronomia - Instituto de Astronomia, 
Geof\'isica e Ci\^encias Atmosf\'ericas, Universidade de 
S\~ao Paulo, S\~ao Paulo, SP 05508-900, Brazil}
\email{vmplacco@astro.iag.usp.br}

\author{Catherine R. Kennedy}
\affil{Department of Physics \& Astronomy and JINA: Joint 
Institute for Nuclear Astrophysics, Michigan State University, 
East Lansing, MI 48824, USA}

\author{Silvia Rossi}
\affil{Departamento de Astronomia - Instituto de Astronomia, 
Geof\'isica e Ci\^encias Atmosf\'ericas, Universidade de 
S\~ao Paulo, S\~ao Paulo, SP 05508-900, Brazil}

\author{Timothy C. Beers, Young Sun Lee}
\affil{Department of Physics \& Astronomy and JINA: Joint Institute for Nuclear Astrophysics, Michigan State University, East Lansing, MI 48824, USA}

\author{Norbert Christlieb}
\affil{Zentrum f\"ur Astronomie der Universit\"at Heidelberg, Landessternwarte, 
K\"onigstuhl 12, 69117, Heidelberg, Germany}

\author{Thirupathi Sivarani}
\affil{Indian Institute of Astrophysics, 2nd block, Koramangala, Bangalore 560034, India}

\author{Dieter Reimers}
\affil{Hamburger Sternwarte, Universit\"at Hamburg, Gojenbergsweg 112, 21029 Hamburg, Germany}

\author{Lutz Wisotzki}
\affil{Astrophysical Institute Potsdam, An der Sternwarte 16, 14482 Potsdam, Germany \\}

\accepted{for publication in AJ -- December 18, 2009}

\begin{abstract}

We have developed a new procedure to search for carbon-enhanced
metal-poor (\cemp{}) stars from the Hamburg/{}ESO (HES) prism-survey
plates. This method employs an extended line index for the CH G-band,
which we demonstrate to have superior behavior when compared to the
narrower G-band index formerly employed to estimate G-band strengths
for these spectra. Although \cemp{} stars have been found previously
among candidate metal-poor stars selected from the HES, the selection
on metallicity undersamples the population of intermediate-metallicity
\cemp{} stars ($-$2.5~$\le$\metal$\le-$1.0); such stars are of
importance for constraining the onset of the s-process in
metal-deficient asymptotic giant-branch stars (thought to be
associated with the origin of carbon for roughly 80\% of \cemp{}
stars). The new candidates also include substantial numbers of warmer
carbon-enhanced stars, which were missed in previous HES searches for
carbon stars due to selection criteria that emphasized stars with
cooler temperatures.

A first subsample, biased towards brighter stars ($B<$ 15.5), has been
extracted from the scanned HES plates. After visual inspection (to
eliminate spectra compromised by plate defects, overlapping spectra,
etc., and to carry out rough spectral classifications), a list of 669
previously unidentified candidate \cemp{} stars was compiled.
Follow-up spectroscopy for a pilot sample of 132 candidates was
obtained with the Goodman spectrograph on the SOAR 4.1m telescope. Our
results show that most of the observed stars lie in the targeted
metallicity range, and possess prominent carbon absorption features at
4300\,{\AA}. The success rate for the identification of new
\cemp{} stars is 43$\%$ (13 out of 30) for \metal~$<-$2.0. 
For stars with \metal~$<-$2.5, the ratio increases to 80$\%$ (4 out of
5 objects), including one star with \metal~$<-$3.0.

\end{abstract}


\section{Introduction}
\label{intro}

The contemporary explosion of information arising from high-resolution
spectroscopic studies of metal-poor stars in the Galaxy is re-shaping
our understanding of the nature of the nucleosynthesis processes that
took place during the early stellar generations. Among the most
interesting are detailed follow-up observations of stars exhibiting large
over-abundances of carbon ($+$0.5~$<$\cfe$<+$4.0), an apparently
common occurance among metal-poor stars \citep{beers2005}.  

It has been reported that a large fraction, at least 20$\%$, of
stars with metallicities \metal\footnote{\abund{A}{B} =
$log(N_A/{}N_B)_{\star} - log(N_A/{}N_B)_{\odot}$, where $N$ is the
number density of atoms of a given element, and the indices refers to
the star ($\star$) and the Sun ($\odot$).}$<$$-$2.0, exhibit large
over-abundances of carbon \citep[\cfe$>$$+$1.0 ;][]{lucatello2006}. The
fraction of so-called carbon-enhanced metal-poor (\cemp) stars rises
to 30$\%$ for \metal$<$$-$3.0, 40$\%$ for \metal$<$$-$3.5, and 100$\%$
for \metal$<$$-$4.0 \citep{christlieb2002,frebel2005,norris2007}. 
However, there are also recent studies \citep[e.g.][]{cohen2005,
frebel2006} claiming that this fraction is somewhat lower (9$\%$ and
14$\%$, respectively). This variety of claims is one of the
motivations for obtaining reliable determinations of metallicities and
carbon abundances for a larger number of stars. Furthermore, the
identification of (in particular, brighter) CEMP stars will play a
major role in theoretical work on the subject \citep{herwig2004,
campbell2008,lau2009}, as they will enable the high-resolution
spectroscopic follow-up required to derive the abundance patterns of
additional elements, and thereby test suggested astrophysical sites
that might be associated with the carbon production. 

The vast majority of known CEMP stars were originally
identified as metal-poor candidates from objective-prism surveys, such
as the \hk{} survey of Beers and colleagues \citep{beers1985,
beers1992}, and the Hamburg/{}ESO Survey \citep[\hes{};
][]{christlieb2003,christlieb2008}, both of which were based on 
the presence of weak (or absent) lines of CaII. A list of \hes{} stars
with strong molecular lines of carbon has been previously published by
\citet{christlieb2001}. Medium-resolution spectra for most of these
objects have been obtained over the past few years \citep{goswami2006,
marsteller2007}. Inspection of these data indicate that at least
50$\%$ of these targets are consistent with identification as \cemp{}
stars, while the others are roughly solar-metallicity carbon-rich
stars. However, this previous set of carbon-rich candidates was
selected based on the \emph{sum} of molecular carbon lines, such as
\cn, \ctwo, and \ch, which over-emphasizes cooler stars in the sample.
\cemp{} stars with effective temperatures higher than about 5500~K
often only exhibit unusual strengths of just a single carbon feature,
the \ch{} G-band at 4300\,{\AA}, and were likely to have been missed in
the previous assembly. Since most previous \cemp{} stars have been
discovered by targeting low-metallicity candidates, this has resulted
in a biasing of the resulting samples of carbon-rich stars to
\metal{}$<$$-$2.5; it would clearly be useful to extend the
metallicity range for their discovery to higher values.

For most CEMP stars there exists a clear correlation between carbon
enhancement and the presence of s-process-element over-abundances,
such as for Ba \citep[\cemp-s stars - see][]{beers2005}. Such behavior
is consistent with the hypothesis that these enhancements (both for
carbon and the s-process elements) are due to nucleosynthesis
processes that took place during the asymptotic giant-branch
\citep[AGB - see] [for a detailed discussion]{herwig2005} stage of
evolution, either from the star itself \citep[which should rarely be
found, but see][]{masseron2006} or by a now-extinct binary companion
that has transferred material to a surviving (observed) component
\citep{stancliffe2008}. 

However, recent studies \citep[e.g.,][]{aoki2007} have shown that this
correlation no longer persists (or at least is different in nature)
for stars with \metal$<$$-$2.7, including all of the most
iron-deficient stars known to date: HE~0107-5240
\citep[~\metal=$-$5.3; ][]{christlieb2004}, HE~1327-2326
\citep[~\metal=$-$5.4; ][]{frebel2005} and HE~0557-4840
\citep[~\metal=$-$4.75; ][]{norris2007}. These so-called \cemp{}-no
stars (indicating a lack of s-process-element over-abundances), and
the other categories of \cemp{} stars that have been noted
\citep{beers2005}, suggest that a variety of mechanisms for the
production of carbon must have played a role in the early Universe.
Furthermore, due to the aforementioned metallicity-dependent selection
bias, many of the \cemp{} stars known to date may be associated with
the outer-halo population, which exhibits a peak metallicity of
\metal{} $\sim -$2.2 \citep{carollo2007}. Additional \cemp{} stars
that are likely to be associated with the inner-halo and metal-weak
thick-disk populations, which extend to higher metallicities, are
required to investigate possible differences in their origins
\citep[e.g.,][]{frebel2006,tumlinson2007}. 

The primary goal of the present work is to demonstrate the efficacy of
searching for intermediate-metallicity \cemp{} stars, through the use
of a new approach for their identification. The inclusion of warmer
carbon-enhanced candidates (which do not exhibit
\cn{} and \ctwo{} bands) also enables investigations between the
observed levels of carbon enhancement and evolutionary stage. It
should also be kept in mind that the inventory of ultra (~\metal$<$$
-$4.0) and hyper (~\metal$<$$-$5.0) metal-poor stars is likely to be
incomplete. Even if some of those extreme objects might not present
carbon enhancements, this work uses the available data as a support to
find candidates that meet our expectations. Such extreme stars may
have been overlooked in previous searches due to noisy spectra in the
region of CaII~K on objective-prism plates \citep[see][for an
alternative procedure to overcome this issue]{christlieb2008}, but
they could reveal themselves by the presence of strong CH G-bands that
are commonly associated with the most iron-deficient stars. 

This paper is outlined as follows. The main features of the \hes{}
stellar database, and its specific application for the present work,
are outlined in \S \ref{database}. Section \ref{index} considers the
flaws of the current line index used by the HES to quantify the
strength of the \ch{} G-band, and provides a definition of a new,
extended line index for the G-band. The first
\hes{} subsample of candidate \cemp{} stars selected on the basis of
this new index, and the criteria for candidate selection, are
discussed in \S \ref{selecand}. Section \ref{followup} reports on
medium-resolution follow-up spectra obtained with the SOAR 4.1m
telescope for 132 \cemp{} candidates in this pilot investigation,
along with determinations of their atmospheric parameters (\teff, 
\logg, \metal{}) and carbon abundances (\cfe). Finally, our
conclusions and perspectives for future observational follow-up are
presented in \S \ref{last}.

\section{The HES Database}
\label{database}

The Hamburg/ESO Survey \citep{reimers1990,reimers1997,wisotzki2000} was
the first all-southern sky quasar survey. The main motivation for the 
survey was to find the brightest quasars in the southern hemisphere, 
both for statistical studies and to identify the best targets for 
follow-up absorption line spectroscopy. Due to the relatively high 
spectral resolution of the ESO Schmidt prism (15\,{\AA} at CaII~K), it
was expected that interesting species of stars, such as
metal-poor halo stars, carbon stars, cataclysmic variables, white
dwarfs, horizontal-branch stars, and others \citep[see][and references
therein]{christlieb2008}, could be found as a byproduct.

The \hes{} prism survey was conducted with the 1m ESO Schmidt
Telecope. With an effective area of 6726 deg$^2$, it covers all the
extragalactic ($|b|>$~30$^o$) southern ($\delta<-$2.5$^o$) portion of
the sky. \citet{christlieb2008} used the survey to increase the number
of metal-poor stars known, compared to the \hk{} Survey, by a factor
of about 3-5, mainly due to the fainter magnitudes achieved
($B$$\sim$17.5). The total survey volume was increased by almost a
factor of 10, relative to the HK survey, but follow-up observations
have not yet been obtained for all of the most interesting \hes{}
candidates. The wavelength coverage of the \hes{} spectra is
3200-5300\,{\AA}, which includes the CaII~K line \citep[3933\,{\AA}, suitable
for \metal{} estimates - see][]{beers1999,rossi2005} and the CH G-band
($\sim$4300\,{\AA}).

The present work (and additional investigations currently in progress)
have made use of the full \hes{} stellar database (4,404,908 objects).
It is most helpful to work with a single and homogeneous sample of
targets, in order to test our new index definitions and still have a
relevant number of candidates for future analysis. Another important
point is that the \hes{} has had a number of published high-resolution
studies that include \cemp{} stars \citep[such as][]{barklem2005,
lucatello2006,aoki2007,schuler2008}, which can be used for comparison.

\section{GPE -- A New Line Index for Carbon}
\label{index}

Previous medium-resolution spectroscopic analyses employed a 15\,{\AA}
wide G-band index (GP), as defined by \citet{beers1999}; a similar
index was originally defined by \citet{beers1985}, prior to the
recognition that such large fractions of metal-poor stars would
exhibit strong carbon enhancement. 

The GP index is a pseudo-equivalent width that measures the contrast between
the observed spectra and the continuum level. It is represented by the 
area enclosed in a 15\,{\AA} wide line-band, delimited by a
psedo-continuum, which is calculated using a linear fit between the
center values of two side-bands, on both blue and red sides of the
line-band. Table \ref{indexlist} lists the wavelength ranges for some
of the G-band indices found in the literature. The need for a new
index is clear, as it has been shown in several studies
\citep[e.g.][]{rossi2005} that the 15\,{\AA} wide line band does not capture
all of the flux absorbed by carbon-related features in the region of
the CH G-band. In addition, the GP index suffers contamination of its sidebands when a given
star is particularly carbon-rich, or at low effective temperatures,
given that a linear fit severely underestimates the level of the continuum for
those objects\footnote{See Figures 1(h) and 2(d) of
\citet{rossi2005}.}.

\begin{center}
\begin{deluxetable}{@{}cccc@{}}[!ht]
\tablecaption{Wavelength Bands (\AA) for CH G-band Indices
\label{indexlist}}
\tablehead{Index & Blue sideband & Line band & Red sideband}
\tablewidth{0pt}
\tabletypesize{\small}
\startdata
GP\tablenotemark{a}	& 4247.0-4267.0 & 4297.5-4312.5 & 4362.0-4372.0  \\
GPHES\tablenotemark{b}	& 4246.0-4255.0 & 4281.0-4307.0 & 4446.0-4612.0  \\
GPE\tablenotemark{c}	& \nodata    	& 4200.0-4400.0 & \nodata    	 
\enddata

\tablenotetext{a}{\citet{beers1999}.}
\tablenotetext{b}{\citet{christlieb2008}.}
\tablenotetext{c}{This work.} 

\end{deluxetable}
\end{center}

\begin{figure}[!ht]
\epsscale{1.15}
\plotone{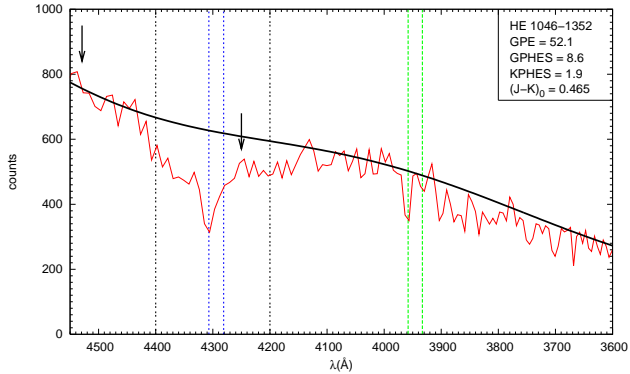}
\caption{Example of the new G-band index definition. The solid (black) line shows the continuum fitting applied to the stars in this work. The long-dashed (green) lines represent the CaII H and K features. Also shown is the comparison between the GPHES (blue dotted lines) and the newly defined \gpe{} (black dashed lines) line bands. The arrows represent the center values of the GPHES continuum sidebands. Note that the wavelength axis is plotted red to blue, as in the original HES scans.}
\label{example}
\end{figure}

\citet{christlieb2008} defined a new G-band index for use with the scanned \hes{}
spectra, \gp{}, of width 26\,{\AA}, calibrated to be on a similar scale as
the GP index. However, as can be appreciated from inspection of Figure
\ref{example}, even this new, wider index is not sufficient for some
of the more extreme \cemp{} candidates identifed in the \hes. A new
index for this particular carbon feature should cover not only the
classical G-band (centered at 4304\,{\AA}), but also the portion of the
spectrum that extends out into the wings of the region, which is
affected by other carbon features (such as \ctwo{}, which are often
exhibited even by warmer \cemp{} stars). Note that even when a star
does not have strong carbon (i.e., exhibits a weak or ``normal''
G-band), such an index should still remain valid, since there will not
be much signal from other features inside the band (except for the
H$_{\gamma}$ Balmer line at 4340\,{\AA}; see below).

The \gpe{} (\gp{} {\it{Extended}}) index is defined as follows:

\begin{equation}
GPE=\int_{4200}^{4400}\left(1-\frac{S(\lambda)}{C(\lambda)}\right)d\lambda
\label{gpeeq}
\end{equation}

\noindent where	$S(\lambda)$ represents the observed spectrum and $C(\lambda)$
is the local continuum. This definition is similar to that of
\citet{cardiel1998}, but here we do not estimate the continuum level by
side-band interpolations, since the presence of the carbon features
can also affect those regions and thereby compromise the index. We
experimented with a variety of fitting approaches for the continuum,
including the same procedures originally adopted for the GP and GPHES
indices. The final choice is based on the techniques employed by the
SEGUE Stellar Parameter Pipeline \citep[SSPP - see][for a detailed
description of the procedure]{lee2008a,lee2008b,allende2008}, adjusted
to work at the resolution of the \hes{} spectra.

Figure \ref{example} shows a typical (cool) carbon-enhanced star
spectrum from the scanned \hes{} plates. The narrow area around
4300\,{\AA} shows the location of the \gphes{} index, which is wider
(and shifted slightly to the blue region) than the GP index. 
The new \gpe{} index line band is represented by the
200\,{\AA} wide region around the same location. Figure \ref{example}
shows that the \gphes{} index band is too narrow to be representative
of the strength of the entire feature, and its sidebands are
contaminated as well. Similar comments apply to the GP index.  

From our own inspection, the optimal definition of the new index
covers the range 4200-4400\,{\AA}. The \gpe{} index does encompass the
H$_{\gamma}$ Balmer line at 4340\,{\AA}, but this should not represent
a problem, since this Balmer line will be present in carbon-normal
stars as well; its strength should scale in the same way with
temperature for both carbon-normal and carbon-rich stars. In the
definition of \gpe{}, the continuum shape plays an especially
important role, since it must be well-fit over the entire region
(rather than estimated from more isolated sidebands).  

\section{Selection of \cemp{} Candidates}
\label{selecand}

The main goal of this pilot study is to test the new \gpe{} index with
the \hes{} database, by comparing its ability to select \cemp{} stars
with available high-resolution analysis \citep[and hence known
atmospheric parameters, and \cfe; e.g.][]{aoki2007} that are similar
to the new stars we seek to identify for future follow-up survey
efforts. We begin by obtaining \gpe{} indices for a selected subsample
of \hes{} candidates, as well as for the \hes{} stars studied by
\citet{aoki2007}, and examine their behavior in a \gpe{} versus \jk{}
diagram, where the near-infrared photometry is taken from \twomass{}
\citep{skrutskie2006}. Because most of the stars will be
``carbon-normal'' (meaning the strength of the \ch{} G-band scales
with metallicity), rather than \cemp{} stars, one can then identify
the locus of stars with enhanced carbon based on their deviation from
the trend associated with carbon-normal stars. 

\subsection{First \hes{} Subsample}
\label{firstsample}

To identify our initial candidates, the following criteria were applied
to the \hes{} database:

\begin{itemize}

\item ph$\_$qual~=~AAA (accurate JHK photometry from \twomass{});
\item objtype~=~stars (removes extended and bright sources);
\item \kp{} $<$ 8.0 (removes stars with clearly too strong CaII~K lines for a 
metal-poor star, regardless of their effective temperature);
\item BHES $<$ 15.5 (bright-object selection, for observations with the SOAR telescope);
\item 0.15 $\leq$ \jk~$\leq$ 0.90 (color range suitable for abundance analysis).

\end{itemize}

\begin{figure}[!ht]
\epsscale{1.15}
\plotone{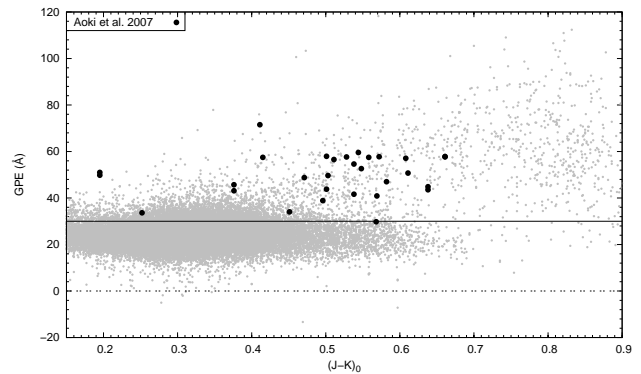}
\caption{Distribution of the new line index, as a function of the \jk{} color, for the 85,894 candidates (small gray dots) and stars from \citet{aoki2007} (black filled circles). The solid line shows the lower limit of \gpe{}.}
\label{combo}
\end{figure}

This first set of constraints yielded 85894 raw candidates. The \gpe{}
index was calculated for those candidates, as well as for 
low-resolution spectra of the \hes{} stars in \citet{aoki2007}, which
are confirmed \cemp{} stars. Figure \ref{combo} shows the distribution
of the \gpe{} index, as a function of \jk{} color, for the raw
candidates (small gray dots) and for the \citet{aoki2007} stars (black
filled circles). The \jk{} color was chosen as a proxy for
temperature, since this variable greatly influences the strength of
molecular carbon features, such as \cn, \ctwo{}, \ch{}, as well as the
hydrogen Balmer lines. The index works because, for a given value of
\jk{}, the \cemp{} stars will have higher \gpe{} values 
than the ones without enhancements.

Based on the location of the known carbon-enhanced stars on this
diagram, relative to the locus of carbon-normal stars, a lower limit
on \gpe{} was set at 30\,{\AA}, reducing the number of candidates to
6018 stars. We are aware that possible candidates may be missed by
this restriction, but this value is a compromise between obtaining a satisfatory
number of candidates to explore for new CEMP stars and the time spent on
the follow-up observations. If the limiting value was chosen at \gpe~=~35, the
yield would be only 1883 candidates. Similarly, going as low as
\gpe~=~25, the number would rise to 26313 candidates.

\begin{figure}[!ht]
\epsscale{1.15}
\plotone{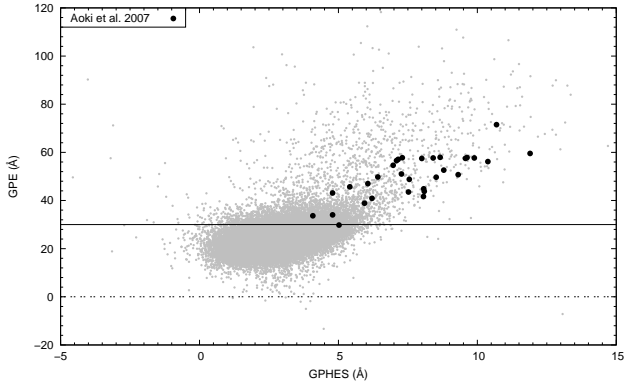}
\caption{Comparison between the \gp{} index previously calculated 
for the \hes{} stars and the new \gpe{} index. The filled black 
circles are stars from \citet{aoki2007}.}
\label{compara}
\end{figure}

One of the primary reasons for the use of a new index is that the
\gphes{} index, as it is calculated for the \hes{} stars, is likely to be
saturated, or have its sidebands contaminated from strong carbon
features. Figure \ref{compara} shows the values for both indices for
the first subsample. One can clearly notice that a deviation from a
linear relation between the two indices occurs, especially for the
higher values of \gpe{} (which are also the stars with redder \jk{},
as seen in Figure \ref{combo}). It is also obvious that the \gpe{}
index enjoys a greater dynamical range than the \gphes{} index, which
is crucial when one considers the effects of errors on the measurement
of these indices. Small measurement errors impact the \gphes{} index
far more than the \gpe{} index is expected to be perturbed (owing to
its larger width and better-defined continuum).

Since the restrictions made above do not take into account any cuts on
S/{}N ratio, nor can they distinguish between satisfactory
measurements of HES spectra and possible difficulties due to plate
artifacts, overlapping spectra, etc., a careful inspection of each
prism spectrum is necessary, as discussed below.

\subsection{Visual Inspection}

\begin{figure}[!ht]
\epsscale{1.15}
\plotone{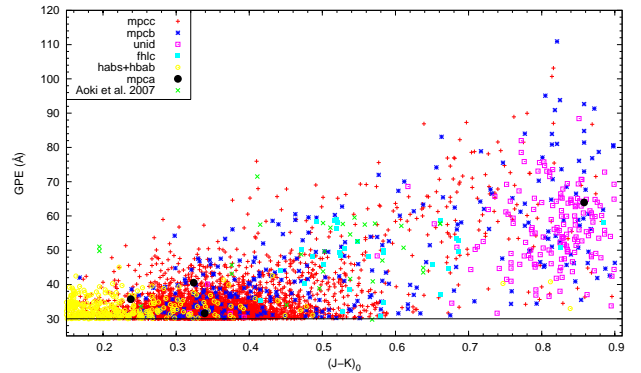}
\caption{Index-color diagram for the 6018 candidates that were subject 
to visual inspection, divided according to the classes defined in Table 
\ref{canclass}. The \citet{aoki2007} stars are indicated by green crosses.}
\label{comp}
\end{figure}

To validate the index calculations, a visual inspection of the
digitized \hes{} spectra was made for candidate stars with \gpe~$\geq$
 30\,{\AA} in order to: (1) assign classifications to the stars based on the
strength of the CaII~K line and the presence of hydrogen Balmer lines,
or clear molecular carbon bands and (2) rule out spurious values of
\gpe{} originating from overlapping spectra, emulsion scratches, or
border effects on the photographic plates. Table \ref{canclass} lists
the distribution of the sample of 6018 candidates, according to the
main assigned classes. The behavior of the new index for all
classes (excluding low S/N spectra and errors due to overlaps 
and artifacts on the photographic plates) is shown in Figure \ref{comp}.    

\begin{center}
\begin{deluxetable}{ccc}[!ht]
\tablecaption{Visual Inspection Classification for the Selected Candidates \label{canclass}}
\tablehead{Tag & Description & Candidates}
\tablewidth{0pt}
\tabletypesize{\small}
\startdata
mpca & Absent CaII~K line                 &    4  \\
mpcb & Weak CaII~K line                   &  280  \\
mpcc & Strong CaII~K line                 & 4614  \\
unid & CaII~K line not found              &  143  \\
fhlc & Faint high latitude carbon stars   &   30  \\
habs & Strong absorption H lines          &   73  \\
hbab & Horizontal-branch/{}$A$ type star  &  218  \\
nois & Low signal-to-noise ratio          &  277  \\
ovl  & Overlapping spectra                &   79  \\
art  & Artifacts on photographic plates   &  123  \\
\enddata
\end{deluxetable}
\end{center}

From inspection of Figure \ref{comp}, the majority of stars on the
blue end of the \jk{} scale exhibit strong hydrogen lines. They came
into the sample due to the fact that the strong H$_{\gamma}$ line
(4340\,{\AA}) contributes significantly to the \gpe{} line index. For
redder colors, \jk{} $>$ 0.3, the strength of the Balmer line
decreases with temperature, so the enhancement of the line index is no
longer a serious issue. On the red end of the color scale, one sees
that the {\it{unid}} stars are concentrated in the \jk{} $>$ 0.7
region. These are cool stars, with little signal in the blue end of
the spectrum, hence it is difficult to identify (and estimate the
strength of) the CaII~K line on the original prism spectra.

\subsection{The Index-Color Selection}

\begin{figure}[!ht]
\epsscale{1.15}
\plotone{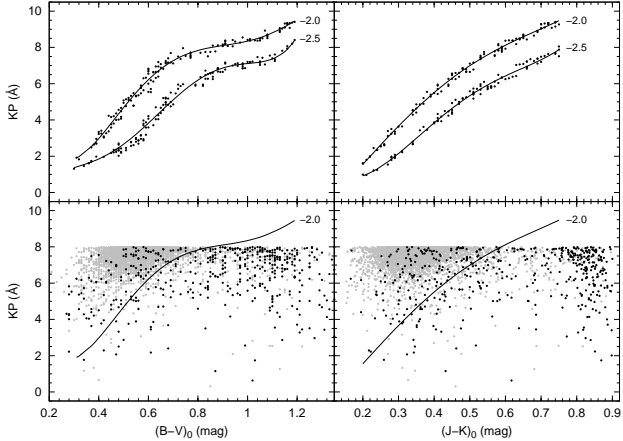}
\caption{Upper panels: Polynomial adjustments for constant values of \metal, 
based on \citet{christlieb2008}. Lower panels: Selection criteria to eliminate 
strong CaII~K line stars ({\emph{mpcc}}). The black dots represent the stars with absent 
({\it{mpca}}), weak ({\it{mpcb}}) or not found ({\it{unid}}~) CaII~K lines. 
The gray dots represent the {\emph{mpcc}} stars and the ones below the 
\metal$\leq$$-$2.0 line for at least one of the color indices are selected.}
\label{kpcutoff}
\end{figure}

One difficulty with the selection described above is that the
candidate sample is dominated by a large number of stars with strong
CaII~K features, many of which may be more metal-rich than the \cemp{}
stars we seek to identify. To reduce the number of these objects, we
adopt a relaxed version of the selection of \citet{christlieb2008} in
the KP\footnote{The KP line index measures the strength of the CaII~K
line, defined by \citet{beers1999}.} index versus \jk{} or BVHES color
parameter space. A metallicity cutoff of \metal{} = $-$2.0, instead of
the \metal{} = $-$2.5 limit used by the HES metal-poor star selection,
was chosen. Note that errors in the measurement of the KP index
ensures that (due to their great numbers) many stars with $-2.0 \le
{\rm \metal{}} \le -1.0$ will still enter our sample. Had we raised
the cutoff in the selection closer to \metal{} $\sim -1.0$, the
numbers of higher-abundance stars would become prohibitive. Figure
\ref{kpcutoff} shows the distribution of stars with strong CaII
K~lines ({\it{mpcc}}) for both colors and the adjusted polynomials.

The final selection of \cemp{} candidates includes all stars with absent 
({\it{mpca}}), weak ({\it{mpcb}}) or not found ({\it{unid}}~) 
CaII~K lines, objects with strong carbon molecular bands ({\it{fhlc}}), 
and also the {\it{mpcc}} stars with KP indices that are
below at least one of the KP cutoffs (gray dots on Figure
\ref{kpcutoff}). After this step, a search was performed on the
full candidate list, and all of the already-known objects (from previous
\hes{} selection, including the metal-poor stars and known
carbon-enhanced stars) were removed. This procedure yielded a list of
669 \cemp{} newly identified candidate \cemp{} stars.

\section{Validation of the \cemp{} Candidates}
\label{followup}

Validation of our selected \cemp{} candidates is an important part of
this pilot study. For this purpose we have obtained medium-resolution
optical spectra for a limited number of \cemp{} candidates with the
SOAR 4.1m telescope. After gathering and reducing the data, we
obtained first-pass estimates of the stellar atmospheric parameters
using the SSPP and a separate procedure to measure \cfe{}. Details of
the observations, reduction procedures, and further analysis are
provided below.

\subsection{Medium-Resolution Spectroscopic Observations}

Medium-resolution spectra for the first 132 of our 669 \cemp{}
candidates were obtained with the new Goodman high-throughput
spectrograph on the SOAR 4.1m telescope, over the course of early
science verification for this instrument. The Goodman spectrograph
operates with several different observing modes. We employed the 600
l/mm grating in the blue setting (wavelength range 3550-5500\,{\AA}) with
a 1.03" slit. This resulted in a resolving power of $R\sim$1500
(resolution of $\sim$3.5\,{\AA}). This resolution was chosen due to its
similarity to spectra obtained during the Sloan Digital Sky Survey
\citep{york2000}, for which the SSPP was designed to work. 

\begin{figure}[!ht]
\epsscale{1.2}
\plotone{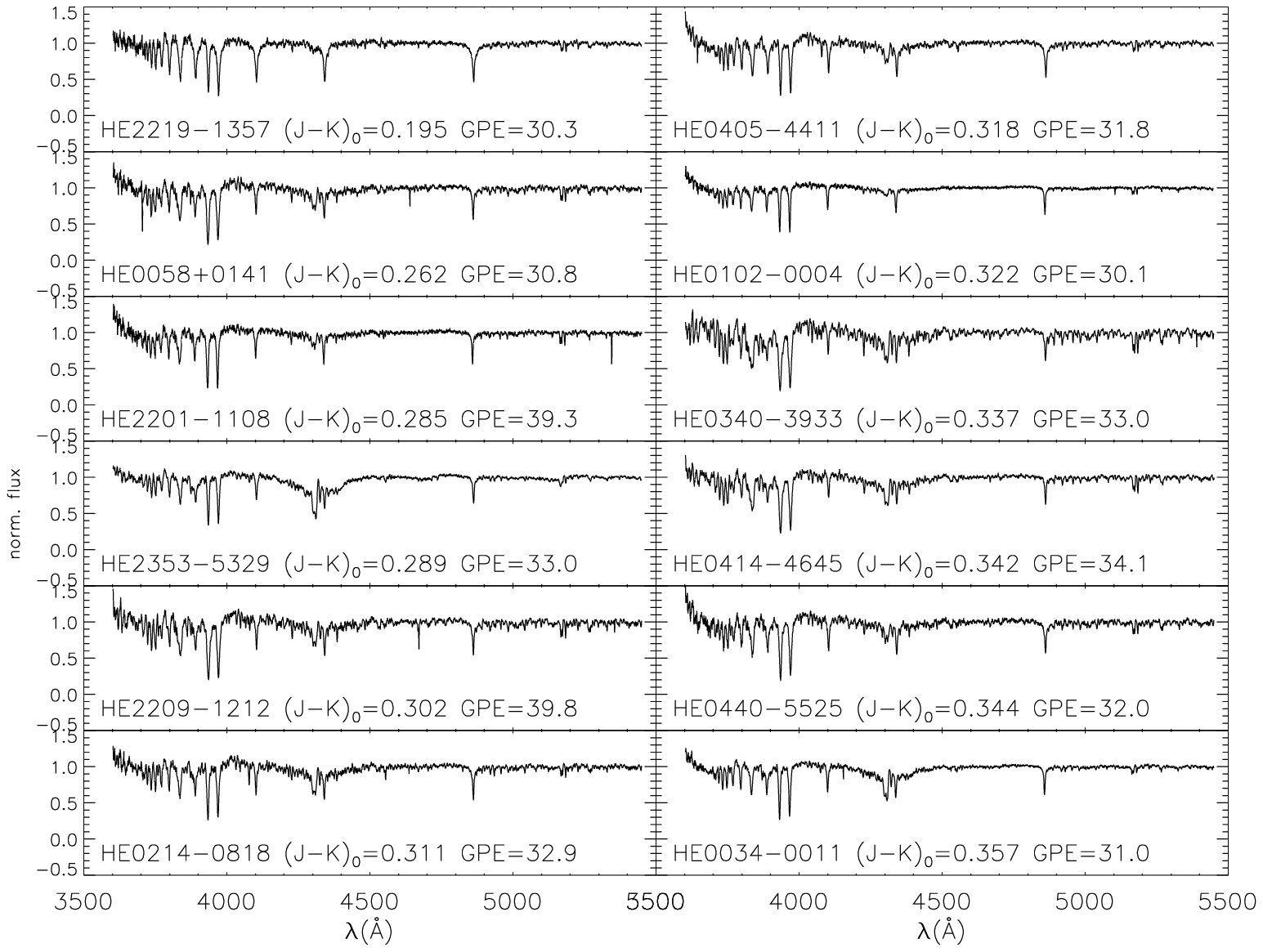}
\caption{Example of \cemp{} candidates observed based on the new 
line index criteria. The spectra were taken with Goodman spectrograph 
on the SOAR telescope.}
\label{spectrajk01}
\end{figure}

\begin{figure}[!ht]
\epsscale{1.2}
\plotone{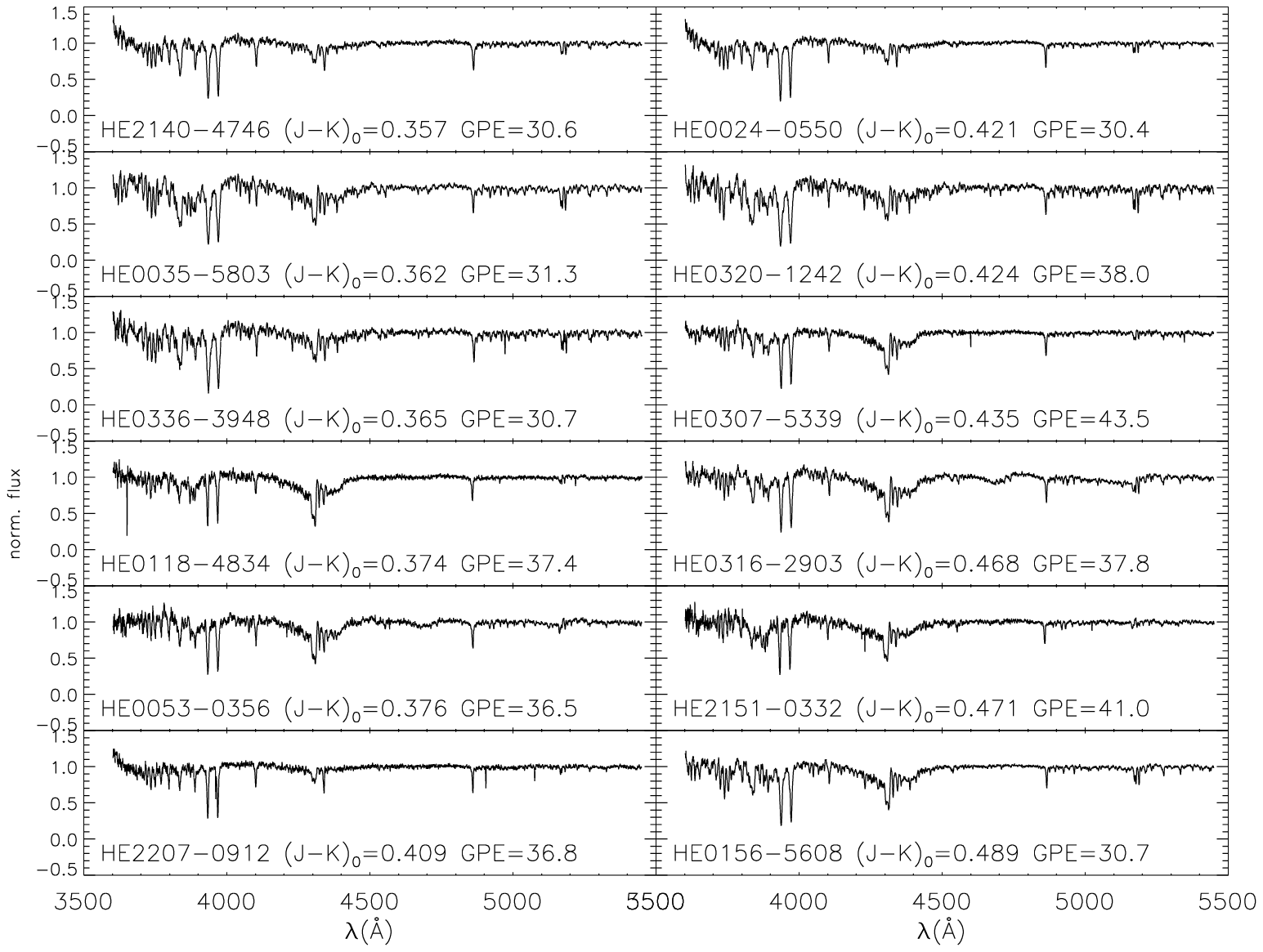}
\caption{Example of \cemp{} candidates observed based on the new 
line index criteria. The spectra were taken with Goodman spectrograph 
on the SOAR telescope.}
\label{spectrajk02}
\end{figure}

\begin{figure}[!ht]
\epsscale{1.2}
\plotone{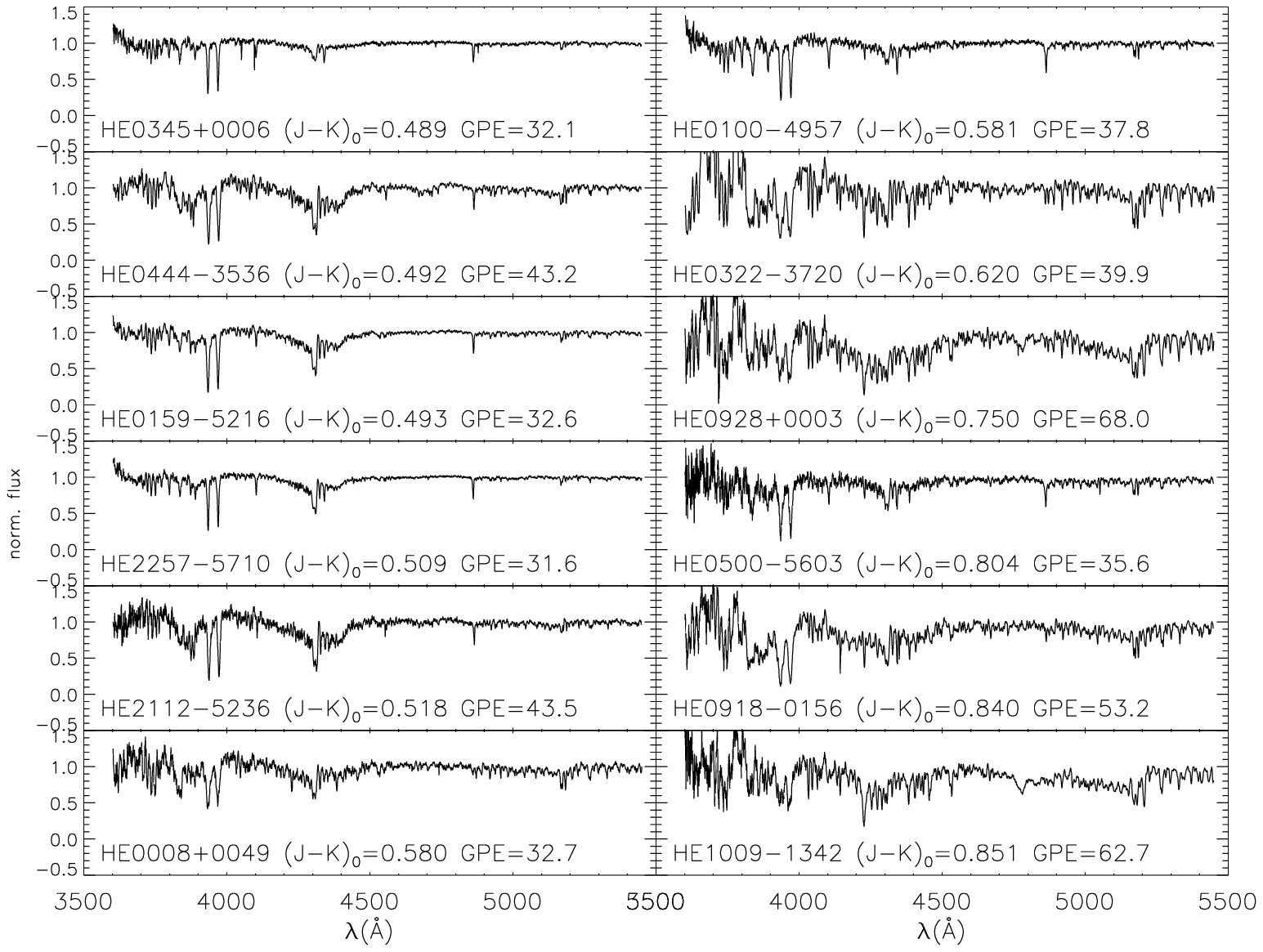}
\caption{Example of \cemp{} candidates observed based on the new 
line index criteria. The spectra were taken with Goodman spectrograph 
on the SOAR telescope.}
\label{spectrajk03}
\end{figure}

The calibration frames include biases, quartz flats, as well as HgAr
and Cu arc lamp exposures taken following each program object's
observation. The exposure times for most of the observed stars were in
the range of 10-20 minutes (targeting an ``as-observed'' S/N $> 40$ in
the region of the CH G-band). Bias subtraction, flat-field correction,
spectral extraction, wavelength calibration, and continuum
normalization were all performed using standard IRAF packages. Table
\ref{candlist} lists the equatorial coordinates, BHES magnitude,
\jk{}, \gpe{}, \kphes{}, \gphes{}, and the classifications for our sample. 
The spectra for some of the observed stars are shown in Figures
\ref{spectrajk01}-\ref{spectrajk03}, organized by increasing \jk{}
color values. 

\subsection{Atmospheric Parameter Estimates and Carbon Abundances}

We employed the SSPP to obtain first-pass atmospheric parameter
estimates for the observed \cemp{} candidates; the results are listed
in Table \ref{atmpar}. The last two columns refer to the carbon
abundance ratios and estimated errors, respectively, obtained by the
procedures discussed below.

The radial velocities calculated for the standard stars in our program
presented unexpectedly large errors (on the order of 50 km/{}s), which
we suspect are due to poorly corrected flexure of the Goodman
spectrograph during commissioning. We used different techniques for
this procedure (including line-by-line estimates and cross-correlation
analysis) to assure that large-than-desired errors were not due to analysis issues. Since
the velocities for the program stars are not known in advance, similar
errors are expected. This does not present a major issue for our
particular application, since the SSPP requires only a rough estimate of
the radial velocity to perform its calculations.  However, it would
clearly to be desirable to improve the derived velocity errors for
future work.  

\begin{center}
\begin{figure}[!ht]
\includegraphics[width=0.8 \columnwidth, angle=270]{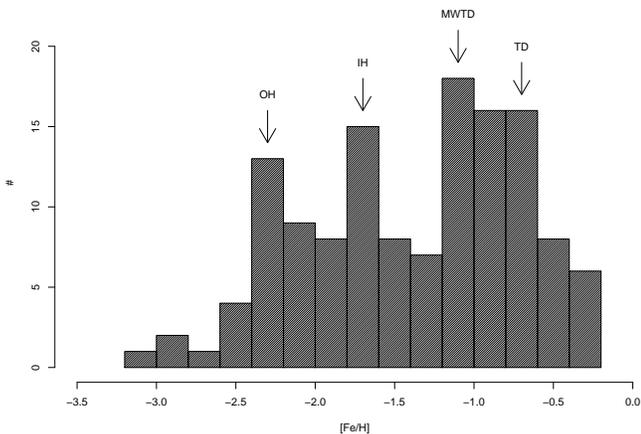}
\caption{Metallicity distribution for the observed candidates. The arrows 
indicate the location of the peak metallicities in the observed distribution, 
which are rather close to those associated by \citet{carollo2007} with the 
outer-halo (OH; \metal{} = $-$2.2), inner-halo (IH; \metal{} = $-$1.6), 
metal-weak thick disk (MWTD; \metal{} = $-$1.3) and cannonical thick-disk 
(TD; \metal{} = $-$0.6) populations. Bins are 0.2 dex in width.}
\label{metald}
\end{figure}
\end{center}

Figure \ref{metald} shows the observed metallicity distribution for
the stars in Table \ref{atmpar}. It is interesting to note that the
two prominent peaks at low metallicity lie rather close to the peak
metallicities that \citet{carollo2007} associate with the outer-halo
(~\metal{} = $-$2.2) and inner-halo (~\metal{} = $-$1.6) populations.
Additional stars that may be associated with the metal-weak and
canonical thick-disk populations are evident at higher metallicity. We
conclude that we are, in fact, obtaining new \cemp{} stars distributed
over our targeted metallicity range.   

\begin{figure}[!ht]
\epsscale{1.15}
\plotone{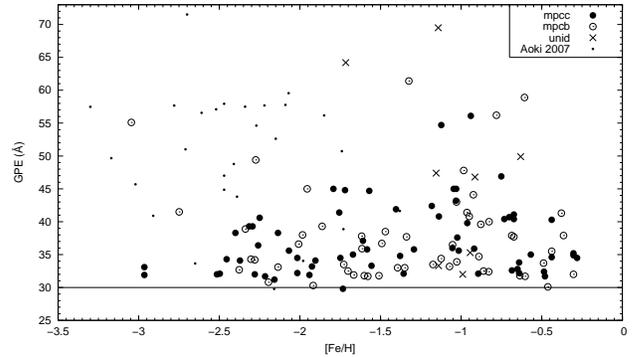}
\caption{Behavior of the metallicity with the \gpe{} index for the 
observed candidates and for the stars from \citet{aoki2007}.}
\label{metaldist}
\end{figure}

Trends of the \gpe{} index with derived metallicity are presented in
Figure \ref{metaldist}. This figure also shows the relationship
between the high-resolution \metal{} obtained by \citet{aoki2007} and
the \gpe{} index, calculated directly from the \hes{} prism spectra.
There is no apparent distinction between the regimes for {\it{mpcb}}
and {\it{mpcc}} stars. Also, as expected, the metallicities greater
than $-$0.5 seen in Figure \ref{metaldist} belong to the stars with
higher temperatures. In fact, one of the main purposes of this work is
to find \cemp{} with metallicities greater than \metal{} = $-$2.5, in
order to fill out the upper-right portion of Figure \ref{metaldist}.

For the estimation of carbon abundances, we generated an extensive 
grid of synthetic spectra covering wavelengths between 3600-4600\,{\AA}. 
The stellar parameters of the grid covers 
\teff{} from 3500 to 9750~K, \logg{} from 0.0 to 5.0 and \metal{} from $-$2.5 to 0.0.  
The carbon abundances (\abund{C}{H}) ranges between
\metal$-$0.5 $\leq$ \abund{C}{H} $\leq$ $+$0.5, for a given value of \metal.
We employed Kurucz NEWODF models \citep{castelli2003} and 
the current version of the spectrum synthesis code \texttt{turbospectrum}
\citep{alvarez-plez} for generating the synthetic spectra.
The linelists used are the same as in \citet{sivarani2006}. 

\begin{figure}[!ht]
\epsscale{1.15}
\plotone{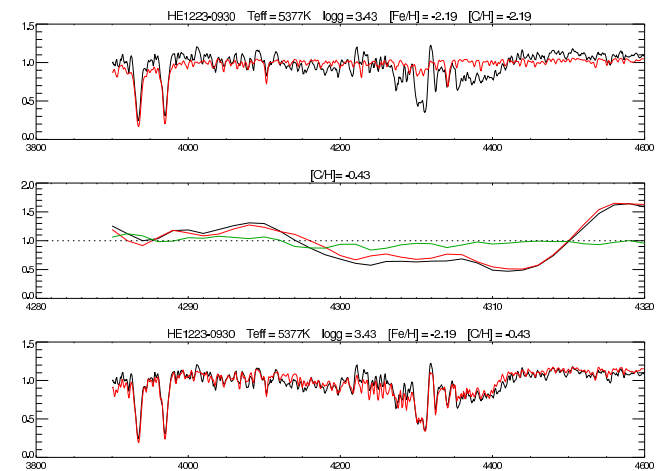}
\caption{Example of carbon abundance determination for one of the stars in our sample. 
The upper panel shows a portion of the original spectrum (black) overlayed 
with a synthetic spectrum (red) with the listed parameters and \cfe~= 0.0. The middle 
panel shows the region around the CH G-band, with a red line showing the best fit. 
The green line is a division of the original spectrum by the fit spectrum, which 
should be close to 1.0 for a successful fit. The lower panel shows the result of 
the best fit with the listed \abund{C}{H}.}
\label{cfesynt}
\end{figure}

Estimation of carbon abundance was accomplished using chi-square
minimization of the observed and synthetic spectra, in the wavelength
region between 4285-4320\,{\AA}. The initial guess value for \abund{C}{H}
was the same as \metal{} (given by the SSPP), i.e., a solar \cfe. An
example fit to the CH G-band region, which is the feature used to
estimate \cfe, is shown in Figure \ref{cfesynt}. Only the carbon
abundance is changed; all the other stellar parameters are kept
constant and chi-square was estimated, using the IDL AMOEBA routine
(down-hill Simplex) for optimization. In most cases the procedure
converged to an adequate fit, by which we mean the chi-square of the
residuals was more than a one-sigma improvement over the initial
(solar) estimate; the typical error bar associated with this situation
is on the order of $\delta_{\rm \cfe}$ = 0.1 dex. In other cases, although
the routine converged, the level of improvement did not reach the
one-sigma level. In these instances we assign errors of 0.2 dex. 
These determinations of \cfe, and their errors, are listed in the
last two columns of Table \ref{atmpar}. To test that the reported
value for \cfe{} is a detection, rather than an upper limit, we
further demand that the integrated line strength in a 20\,{\AA} band
(from 4295\,{\AA} to 4315\,{\AA}) be at least 1.5\,{\AA}. This value was settle
upon by comparison with noise-injected synthetic spectra with a
variety of input fixed \cfe. Determinations of \cfe{} that failed to
meet this criterion are considered upper limits, and are reported in
Table \ref{atmpar} without listed errors. 

\begin{figure}[!ht]
\epsscale{1.2}
\plotone{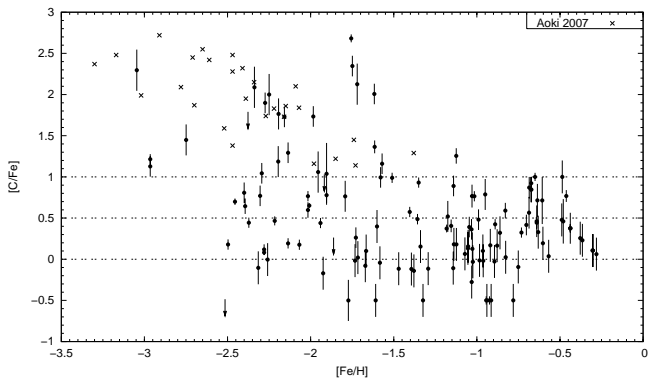}
\caption{Behavior of the metallicity with the carbon abundance \abund{C}{Fe} for the observed candidates and for the stars from \citet{aoki2007}. The arrows represent upper limits. The dashed lines show constant values of \cfe{} (0.0, $+$0.5 and $+$1.0).}
\label{carbondist}
\end{figure}

Figure \ref{carbondist} shows the behavior of the carbon abundances
as a function of metallicity, including the high-resolution
measurements from \citet{aoki2007}. As already pointed out by other
studies \citep{rossi2005,lucatello2006}, there is a clear trend in the
\cfe{} ratios, which are higher for lower metallicities, and exhibit 
increasing scatter for \metal~$<-$2.0.  This behavior is seen as well
for the high-resolution data shown in the figure. 

By using our new selection method, which it should be recalled is 
biased towards finding stars with higher carbon
abundance, the fraction of carbon-enhanced stars (considering the
error bars in \cfe), is $\sim$25$\%$. If one considers only the very metal-poor
stars (\metal$<$$-$2.0), the fraction increases to 43$\%$. For the
five observed candidates that present estimated metallicities
significantly lower than \metal{} = $-$2.5, our method reached 80$\%$
success. It is worth noting that the majority of metal-poor stars in
our candidate pool with \metal~$<-$1.0 (51$\%$) present considerable
carbon enhancements (\cfe~$>+$0.5).

\section{Conclusions}
\label{last}

We have developed a new line index for the region of the carbon G-band
at 4304\,{\AA}, \gpe, which has the advantage of capturing more
information concerning the abundance of carbon, since its width takes
into account the wings of the band, which includes other nearby carbon
features. Furthermore, it is not subject to confounding (as were previously employed
narrower indices) due to sidebands that fall in regions of the spectrum
for which carbon features are present. To test this new method, we
obtained a sample of stars from the \hes{} stellar database, and
compared the newly calculated index with the ones for confirmed
carbon-rich stars based on high-resolution analysis \citep{aoki2007}.
Medium-resolution spectra for a sample of 132 stars selected by this
procedure have been obtained with the Goodman spectrograph on the SOAR
4.1m telescope. Our new selection technique achieves a success rate
for newly identified \cemp{} stars of 43\% for stars with \metal~$<-$2.0; 
four out of five 
candidates with \metal~$<-$2.5 exhibit high carbon enhancements
\cfe~$>+$1.0.  It should be kept in mind that these values are not
unbiased estimates of the fractions of CEMP stars, rather, they
indicate the efficacy of our new approach for the identification of
likely carbon-enhanced stars.

We plan to continue our survey for unrecognized \cemp{} stars, based
on this new selection scheme, with the goal of reaching a total sample
of $\sim$1000 such stars. In the past, \cemp{} stars were either
selected as (1) candidate metal-poor stars from the HK survey or
\hes{} based on the apparent weakness of their CaII~K lines (and then
later found to be \cemp{} stars based on medium-resolution
spectroscopic follow-up), or (2) were selected as carbon-rich stars on
the basis of the \emph{sum} of various carbon features in their prism
spectra \citep{christlieb2001}. Both of these techniques have limitations. Technique
(1) clearly misses warmer \cemp{} stars with metallicity 
\metal{}~$>-$2.5, and (due to the color range used in the selection) misses
\cemp{} stars with estimated $B-V >$ 0.9, as the presence of strong
lines of carbon ``reddens'' the inferred colors outside of the
selection window. Technique (2) identifies mostly very cool
carbon-rich stars, since it targets a threshold for the total strength
of carbon features in a stellar spectrum. Even stars with quite strong
\ch{} G-bands often fail to meet the selection threshold, if they are
warm enough to not exhibit
\cn{} and \ctwo{} bands. 

The expanded list of CEMP stars we seek to identify will enable more 
detailed studies at high spectral resolution, in order to assign them 
into their proper sub-classes, and to determine the full set of elemental 
abundances needed in order to explore the astrophysical sites 
associated with the carbon production.

\acknowledgments V.M.P. acknowledges hospitality at the Zentrum f\"ur Astronomie der
Universit\"at Heidelberg, Landessternwarte, during which the visual
inspection of the new \cemp{} candidates took place. V.M.P. and S.R.
acknowledge CNPq, CAPES (PROEX), FAPESP funding (2007/{}04356-3) and
JINA. C.R.K., T.C.B., Y.S.L. and T.S. acknowledge partial support for this
work from grants AST 07-07776, PHY 02-15783 and PHY 08-22648; Physics
Frontier Center/{}Joint Institute or Nuclear Astrophysics (JINA),
awarded by the US National Science Foundation.

\clearpage
\LongTables

\begin{deluxetable}{@{}cccccccccc@{}}
\tablewidth{0pt}
\tabletypesize{\small}
\tablecaption{Stellar Data for the Observed Candidates \label{candlist}}
\tablehead{Name & $\alpha$(J2000) & $\delta$(J2000) & BHES & \jk{} & \gpe{} (\AA) & \kphes{} (\AA) & \gphes{} (\AA) & Tag}
\startdata

HE~0008$+$0049 & 00:11:10.5 & $+$01:05:51 & 14.5 & 0.58 & 32.0 & 7.2 & 4.7 & mpcb \\
HE~0024$-$0550 & 00:26:33.7 & $-$05:33:35 & 14.7 & 0.42 & 30.4 & 6.0 & 3.2 & mpcc \\
HE~0034$-$0011 & 00:36:51.2 & $+$00:05:29 & 15.0 & 0.36 & 29.7 & 5.5 & 5.2 & mpcc \\
HE~0035$-$5803 & 00:37:27.3 & $-$57:47:27 & 15.0 & 0.36 & 31.3 & 7.4 & 4.1 & mpcc \\
HE~0053$-$0356 & 00:56:04.7 & $-$03:40:40 & 14.7 & 0.38 & 36.5 & 6.1 & 6.1 & mpcb \\
HE~0058$+$0141 & 01:01:16.5 & $+$01:57:46 & 15.0 & 0.26 & 28.6 & 6.4 & 2.4 & mpcb \\
HE~0100$-$4957 & 01:02:13.8 & $-$49:41:29 & 15.0 & 0.58 & 37.8 & 7.7 & 2.8 & mpcc \\
HE~0102$-$0004 & 01:05:09.8 & $+$00:11:38 & 14.3 & 0.32 & 29.3 & 6.1 & 3.9 & mpcb \\
HE~0118$-$4834 & 01:20:18.4 & $-$48:19:12 & 14.7 & 0.37 & 37.4 & 5.4 & 5.5 & mpcb \\
HE~0156$-$5608 & 01:58:38.8 & $-$55:54:25 & 14.9 & 0.49 & 30.7 & 6.8 & 5.0 & mpcc \\
HE~0159$-$5216 & 02:01:40.6 & $-$52:02:15 & 14.7 & 0.49 & 32.6 & 7.5 & 5.4 & mpcc \\
HE~0214$-$0818 & 02:16:44.1 & $-$08:04:31 & 14.8 & 0.31 & 32.9 & 6.8 & 4.0 & mpcb \\
HE~0307$-$5339 & 03:08:42.2 & $-$53:28:20 & 14.9 & 0.44 & 43.5 & 7.3 & 7.6 & mpcb \\
HE~0316$-$2903 & 03:18:14.7 & $-$28:52:51 & 14.7 & 0.47 & 37.8 & 6.7 & 5.9 & mpcc \\
HE~0320$-$1242 & 03:23:07.3 & $-$12:31:27 & 15.0 & 0.42 & 38.1 & 6.9 & 3.9 & mpcb \\
HE~0322$-$3720 & 03:24:27.8 & $-$37:09:57 & 14.2 & 0.62 & 39.9 & 7.9 & 5.5 & mpcb \\
HE~0336$-$3948 & 03:38:43.3 & $-$39:38:22 & 14.9 & 0.37 & 30.7 & 6.0 & 5.0 & mpcc \\
HE~0340$-$3933 & 03:41:56.5 & $-$39:24:06 & 14.6 & 0.34 & 33.0 & 6.6 & 4.0 & mpcc \\
HE~0345$+$0006 & 03:48:19.4 & $+$00:15:10 & 15.1 & 0.53 & 30.6 & 6.6 & 4.0 & mpcc \\
HE~0405$-$4411 & 04:07:14.2 & $-$44:03:53 & 15.1 & 0.32 & 31.8 & 6.9 & 1.7 & unid \\
HE~0414$-$4645 & 04:16:10.2 & $-$46:38:17 & 15.1 & 0.34 & 34.1 & 5.7 & 3.4 & mpcc \\
HE~0440$-$5525 & 04:42:00.1 & $-$55:19:30 & 15.0 & 0.34 & 32.0 & 6.4 & 4.1 & mpcb \\
HE~0444$-$3536 & 04:46:39.5 & $-$35:31:07 & 14.7 & 0.49 & 43.2 & 7.7 & 6.3 & mpcc \\
HE~0449$-$1617 & 04:52:01.4 & $-$16:12:11 & 15.1 & 0.42 & 31.7 & 6.7 & 3.8 & mpcb \\
HE~0451$-$3127 & 04:53:45.5 & $-$31:22:18 & 15.1 & 0.50 & 30.4 & 6.6 & 4.4 & mpcc \\
HE~0500$-$5603 & 05:01:41.2 & $-$55:58:46 & 14.7 & 0.80 & 35.6 & 7.9 & 5.8 & mpcc \\
HE~0509$-$1611 & 05:11:30.0 & $-$16:07:43 & 15.1 & 0.52 & 41.7 & 7.8 & 7.1 & mpcc \\
HE~0511$-$3411 & 05:13:40.7 & $-$34:08:16 & 15.0 & 0.37 & 33.1 & 6.4 & 4.5 & mpcc \\
HE~0514$-$5449 & 05:15:11.9 & $-$54:46:21 & 15.0 & 0.31 & 31.0 & 6.3 & 3.2 & mpcb \\
HE~0518$-$3941 & 05:20:23.1 & $-$39:38:18 & 14.6 & 0.18 & 30.9 & 6.0 & 2.5 & mpcc \\
HE~0535$-$4842 & 05:36:51.6 & $-$48:40:50 & 14.7 & 0.39 & 30.5 & 7.8 & 5.1 & unid \\
HE~0536$-$5647 & 05:37:18.1 & $-$56:46:08 & 14.1 & 0.49 & 31.5 & 7.4 & 4.1 & mpcb \\
HE~0537$-$4849 & 05:38:39.1 & $-$48:47:36 & 14.9 & 0.39 & 30.5 & 7.8 & 4.0 & mpcb \\
HE~0901$-$0003 & 09:03:53.6 & $+$00:15:48 & 15.1 & 0.43 & 31.1 & 7.5 & 4.6 & mpcc \\
HE~0910$-$0126 & 09:13:26.1 & $-$01:39:19 & 14.8 & 0.26 & 28.8 & 4.5 & 2.9 & mpcb \\
HE~0912$+$0200 & 09:15:30.1 & $+$01:47:29 & 15.1 & 0.50 & 45.4 & 7.6 & 8.9 & mpcc \\
HE~0918$-$0156 & 09:21:06.2 & $-$02:08:58 & 15.1 & 0.84 & 53.2 & 7.9 & 7.7 & mpcc \\
HE~0922$-$0337 & 09:25:15.3 & $-$03:50:36 & 14.7 & 0.61 & 33.3 & 8.0 & 5.1 & mpcc \\
HE~0923$-$0323 & 09:26:00.7 & $-$03:36:57 & 15.1 & 0.39 & 30.2 & 7.8 & 5.0 & mpcc \\
HE~0928$+$0003 & 09:30:33.2 & $+$00:10:08 & 14.9 & 0.75 & 68.0 & 7.3 & 8.0 & unid \\
HE~0928$+$0059 & 09:31:07.0 & $+$00:46:43 & 14.8 & 0.27 & 30.9 & 7.5 & 4.4 & mpcb \\
HE~0933$-$0733 & 09:36:09.5 & $-$07:46:57 & 15.1 & 0.38 & 41.5 & 7.8 & 6.0 & mpcb \\
HE~0934$-$1058 & 09:36:33.7 & $-$11:11:42 & 14.9 & 0.67 & 44.8 & 7.9 & 6.1 & fhlc \\
HE~0948$+$0107 & 09:51:27.8 & $+$00:53:21 & 14.9 & 0.50 & 31.6 & 5.8 & 3.6 & mpcb \\
HE~0948$-$0234 & 09:51:09.5 & $-$02:48:21 & 15.1 & 0.37 & 34.0 & 7.6 & 4.4 & mpcb \\
HE~0950$-$0401 & 09:52:43.7 & $-$04:16:03 & 14.1 & 0.34 & 36.3 & 6.4 & 5.7 & mpcb \\
HE~0950$-$1248 & 09:53:04.3 & $-$13:03:07 & 15.0 & 0.38 & 33.7 & 7.0 & 4.8 & mpcc \\
HE~0951$+$0114 & 09:53:55.5 & $+$01:00:29 & 14.9 & 0.63 & 59.9 & 7.8 & 5.5 & mpcb \\
HE~1001$-$1621 & 10:03:54.8 & $-$16:35:45 & 15.0 & 0.40 & 34.4 & 6.2 & 4.5 & mpcc \\
HE~1002$-$1405 & 10:04:35.4 & $-$14:19:54 & 14.1 & 0.36 & 38.8 & 7.5 & 4.9 & mpcc \\
HE~1007$-$1524 & 10:09:38.2 & $-$15:39:20 & 15.0 & 0.36 & 32.3 & 6.9 & 4.6 & mpcc \\
HE~1009$-$1342 & 10:12:10.0 & $-$13:57:17 & 15.0 & 0.85 & 62.7 & 8.0 & 5.5 & unid \\
HE~1009$-$1613 & 10:11:26.5 & $-$16:28:40 & 14.4 & 0.40 & 39.6 & 7.0 & 6.9 & mpcc \\
HE~1009$-$1646 & 10:12:11.5 & $-$17:01:17 & 15.1 & 0.40 & 39.2 & 6.6 & 7.5 & mpcc \\
HE~1010$-$1445 & 10:13:03.8 & $-$15:00:51 & 15.0 & 0.56 & 30.6 & 6.9 & 5.8 & mpcc \\
HE~1022$-$0730 & 10:24:39.3 & $-$07:45:59 & 14.9 & 0.37 & 30.2 & 7.7 & 5.3 & mpcb \\
HE~1027$-$1217 & 10:29:29.9 & $-$12:32:31 & 15.1 & 0.43 & 35.2 & 5.4 & 3.1 & mpcb \\
HE~1028$-$1505 & 10:31:23.4 & $-$15:20:46 & 15.0 & 0.62 & 33.5 & 7.8 & 4.4 & mpcc \\
HE~1039$-$1019 & 10:42:25.4 & $-$10:34:51 & 14.9 & 0.40 & 36.2 & 7.8 & 4.8 & mpcb \\
HE~1045$+$0226 & 10:48:03.4 & $+$02:10:47 & 15.0 & 0.57 & 53.6 & 7.4 & 8.9 & mpcb \\
HE~1046$-$1644 & 10:49:13.4 & $-$17:00:19 & 14.7 & 0.55 & 30.2 & 7.0 & 4.4 & mpcb \\
HE~1049$-$0922 & 10:52:26.2 & $-$09:38:33 & 14.7 & 0.58 & 48.4 & 8.0 & 5.5 & unid \\
HE~1049$-$1025 & 10:51:44.2 & $-$10:41:05 & 14.1 & 0.45 & 54.7 & 7.4 & 9.6 & mpcb \\
HE~1104$-$0238 & 11:07:00.4 & $-$02:54:17 & 15.0 & 0.90 & 33.8 & 7.9 & 5.5 & unid \\
HE~1110$-$1625 & 11:13:05.4 & $-$16:41:29 & 15.0 & 0.38 & 33.4 & 6.9 & 5.4 & mpcc \\
HE~1112$-$0203 & 11:14:48.6 & $-$02:19:26 & 14.2 & 0.83 & 45.3 & 7.9 & 5.9 & unid \\
HE~1125$-$1343 & 11:28:26.1 & $-$13:59:58 & 15.0 & 0.66 & 36.1 & 7.1 & 5.4 & mpcc \\
HE~1129$-$1405 & 11:32:19.2 & $-$14:21:44 & 15.1 & 0.48 & 33.0 & 6.5 & 4.9 & mpcc \\
HE~1132$-$0915 & 11:35:24.9 & $-$09:32:33 & 14.7 & 0.39 & 31.8 & 4.8 & 3.4 & mpcc \\
HE~1133$-$0802 & 11:35:59.0 & $-$08:18:43 & 14.9 & 0.49 & 40.4 & 8.0 & 7.1 & mpcc \\
HE~1135$-$0800 & 11:38:23.9 & $-$08:16:57 & 15.1 & 0.54 & 32.7 & 6.0 & 4.9 & mpcb \\
HE~1137$-$1259 & 11:39:37.2 & $-$13:15:52 & 15.0 & 0.58 & 35.0 & 7.3 & 5.3 & mpcb \\
HE~1142$-$0637 & 11:45:00.8 & $-$06:54:18 & 14.9 & 0.57 & 34.3 & 7.4 & 3.8 & mpcc \\
HE~1146$-$1040 & 11:49:24.5 & $-$10:56:41 & 15.0 & 0.50 & 40.9 & 7.6 & 5.4 & mpcc \\
HE~1146$-$1126 & 11:49:09.5 & $-$11:43:02 & 14.9 & 0.58 & 34.9 & 6.5 & 5.3 & mpcc \\
HE~1147$-$1057 & 11:49:33.0 & $-$11:14:26 & 15.1 & 0.38 & 33.2 & 5.7 & 4.4 & mpcb \\
HE~1148$-$1020 & 11:51:11.4 & $-$10:37:32 & 15.0 & 0.41 & 36.2 & 7.7 & 3.6 & mpcb \\
HE~1148$-$1025 & 11:50:49.8 & $-$10:41:42 & 14.8 & 0.42 & 38.5 & 6.9 & 6.1 & mpcb \\
HE~1212$-$1123 & 12:14:36.7 & $-$11:39:48 & 15.1 & 0.29 & 31.5 & 6.2 & 3.9 & mpcb \\
HE~1217$-$1054 & 12:19:56.9 & $-$11:11:27 & 14.9 & 0.55 & 38.3 & 7.6 & 6.0 & mpcc \\
HE~1217$-$1633 & 12:20:30.2 & $-$16:49:44 & 14.8 & 0.52 & 56.5 & 7.7 & 9.7 & fhlc \\
HE~1222$-$1631 & 12:24:59.5 & $-$16:48:15 & 14.8 & 0.57 & 34.1 & 7.7 & 5.7 & mpcc \\
HE~1223$-$0930 & 12:26:01.9 & $-$09:47:35 & 14.5 & 0.50 & 45.8 & 7.4 & 8.2 & fhlc \\
HE~1224$-$0723 & 12:27:15.1 & $-$07:40:21 & 14.9 & 0.41 & 38.9 & 6.7 & 5.1 & mpcc \\
HE~1224$-$1043 & 12:26:51.5 & $-$11:00:35 & 14.8 & 0.36 & 33.5 & 4.8 & 2.5 & mpcc \\
HE~1228$-$0750 & 12:31:30.3 & $-$08:06:38 & 15.0 & 0.30 & 30.3 & 4.8 & 1.9 & mpcb \\
HE~1228$-$1438 & 12:30:44.6 & $-$14:55:05 & 14.5 & 0.89 & 42.6 & 8.0 & 7.9 & mpcb \\
HE~1231$-$3136 & 12:34:31.2 & $-$31:52:39 & 15.1 & 0.33 & 30.3 & 5.6 & 2.2 & mpcb \\
HE~1255$-$2734 & 12:58:18.4 & $-$27:50:23 & 14.3 & 0.43 & 36.8 & 5.6 & 5.9 & mpcc \\
HE~1301$+$0014 & 13:03:45.8 & $+$00:01:28 & 15.1 & 0.46 & 32.6 & 5.2 & 3.3 & mpcc \\
HE~1301$-$1405 & 13:04:03.6 & $-$14:21:30 & 15.1 & 0.48 & 34.3 & 6.7 & 4.8 & mpcc \\
HE~1302$-$0954 & 13:04:58.2 & $-$10:10:11 & 14.5 & 0.49 & 32.8 & 7.4 & 5.1 & mpcb \\
HE~1311$-$3002 & 13:13:59.7 & $-$30:18:21 & 14.3 & 0.58 & 34.8 & 8.0 & 7.2 & fhlc \\
HE~1320$-$1130 & 13:23:37.0 & $-$11:46:03 & 15.1 & 0.34 & 34.4 & 7.0 & 4.7 & mpcb \\
HE~1320$-$1641 & 13:23:11.9 & $-$16:56:38 & 15.0 & 0.87 & 43.5 & 7.9 & 6.9 & mpcc \\
HE~1321$-$1652 & 13:24:27.3 & $-$17:07:48 & 15.0 & 0.35 & 43.3 & 5.8 & 7.1 & mpcc \\
HE~1343$+$0137 & 13:46:17.3 & $+$01:22:29 & 15.1 & 0.41 & 28.3 & 5.5 & 2.5 & mpcc \\
HE~1408$-$0444 & 14:10:50.4 & $-$04:58:51 & 14.7 & 0.22 & 31.2 & 2.2 & 2.8 & mpcb \\
HE~1409$-$1134 & 14:11:43.4 & $-$11:49:02 & 15.0 & 0.36 & 36.4 & 7.9 & 4.3 & mpcb \\
HE~1410$-$0549 & 14:13:21.7 & $-$06:03:33 & 14.9 & 0.25 & 31.0 & 6.1 & 1.4 & mpcb \\
HE~1414$-$1644 & 14:17:03.4 & $-$16:58:23 & 14.6 & 0.47 & 32.8 & 6.0 & 4.7 & mpcc \\
HE~1418$-$1634 & 14:20:51.0 & $-$16:47:46 & 15.1 & 0.54 & 30.5 & 7.0 & 6.1 & mpcc \\
HE~1428$-$0851 & 14:30:40.6 & $-$09:05:09 & 14.9 & 0.53 & 30.5 & 6.2 & 2.4 & mpcc \\
HE~1430$-$1518 & 14:32:56.4 & $-$15:31:35 & 14.9 & 0.79 & 45.9 & 7.6 & 10.1 & unid \\
HE~1447$-$1533 & 14:49:54.5 & $-$15:46:22 & 14.3 & 0.83 & 34.5 & 7.9 & 6.9 & mpcc \\
HE~1448$-$1406 & 14:50:53.1 & $-$14:19:14 & 14.9 & 0.37 & 30.4 & 5.8 & 2.7 & mpcb \\
HE~1451$-$0659 & 14:54:03.0 & $-$07:11:40 & 14.5 & 0.63 & 37.0 & 7.9 & 5.6 & mpcb \\
HE~1458$-$0923 & 15:00:45.4 & $-$09:35:49 & 14.4 & 0.41 & 47.9 & 6.6 & 6.9 & mpcb \\
HE~1458$-$1022 & 15:01:35.7 & $-$10:33:54 & 14.7 & 0.54 & 30.2 & 7.0 & 5.3 & mpcc \\
HE~1458$-$1226 & 15:01:32.8 & $-$12:37:57 & 15.1 & 0.47 & 43.5 & 7.4 & 6.9 & mpcc \\
HE~1504$-$1534 & 15:07:46.2 & $-$15:45:31 & 14.8 & 0.85 & 38.9 & 8.0 & 7.3 & mpcc \\
HE~1505$-$0826 & 15:08:04.7 & $-$08:38:22 & 14.9 & 0.25 & 32.2 & 7.5 & 3.6 & mpcb \\
HE~1507$-$1055 & 15:10:09.9 & $-$11:07:19 & 14.9 & 0.80 & 39.3 & 7.8 & 8.8 & mpcc \\
HE~1507$-$1104 & 15:09:45.4 & $-$11:16:09 & 15.1 & 0.90 & 46.3 & 7.3 & 8.3 & mpcb \\
HE~1512$+$0149 & 15:15:08.3 & $+$01:38:05 & 15.0 & 0.67 & 54.6 & 7.1 & 8.7 & mpcc \\
HE~1516$-$0107 & 15:18:54.0 & $-$01:18:50 & 15.0 & 0.43 & 35.1 & 5.0 & 4.5 & mpcb \\
HE~1518$-$0541 & 15:21:20.6 & $-$05:52:08 & 14.1 & 0.54 & 32.4 & 6.8 & 3.5 & mpcb \\
HE~1527$-$0740 & 15:30:18.5 & $-$07:50:50 & 15.1 & 0.44 & 37.8 & 6.2 & 1.7 & mpcb \\
HE~1529$-$0838 & 15:31:54.8 & $-$08:48:39 & 15.1 & 0.38 & 36.4 & 7.9 & 4.9 & mpcb \\
HE~2025$-$5221 & 20:29:38.6 & $-$52:11:22 & 14.8 & 0.39 & 39.1 & 4.7 & 6.1 & mpcc \\
HE~2052$-$5610 & 20:56:34.9 & $-$55:59:17 & 15.0 & 0.27 & 39.9 & 6.0 & 7.0 & mpcc \\
HE~2112$-$5236 & 21:16:09.2 & $-$52:23:30 & 14.8 & 0.52 & 43.5 & 7.1 & 7.0 & mpcc \\
HE~2117$-$6018 & 21:21:26.2 & $-$60:05:33 & 15.0 & 0.59 & 31.7 & 6.6 & 4.9 & mpcc \\
HE~2140$-$4746 & 21:44:06.1 & $-$47:32:59 & 14.7 & 0.36 & 30.6 & 5.8 & 3.1 & mpcc \\
HE~2151$-$0332 & 21:53:58.6 & $-$03:18:09 & 15.0 & 0.47 & 40.0 & 5.7 & 5.1 & mpcb \\
HE~2201$-$1108 & 22:04:08.4 & $-$10:53:33 & 15.0 & 0.29 & 39.3 & 6.0 & 4.2 & mpcb \\
HE~2207$-$0912 & 22:10:13.4 & $-$08:57:29 & 15.0 & 0.41 & 36.8 & 4.3 & 1.5 & mpcc \\
HE~2209$-$1212 & 22:11:44.1 & $-$11:57:37 & 14.6 & 0.30 & 39.8 & 5.1 & 4.3 & mpcb \\
HE~2219$-$1357 & 22:22:28.2 & $-$13:42:06 & 14.9 & 0.20 & 30.3 & 4.4 & 2.3 & mpcb \\
HE~2231$-$0710 & 22:33:56.1 & $-$06:54:35 & 14.6 & 0.43 & 57.4 & 1.2 & 7.7 & mpcb \\
HE~2257$-$5710 & 23:00:40.4 & $-$56:54:15 & 14.7 & 0.51 & 31.6 & 6.6 & 5.3 & mpcc \\
HE~2353$-$5329 & 23:55:49.3 & $-$53:12:39 & 13.9 & 0.29 & 33.0 & 4.6 & 4.1 & mpcc \\

\enddata
\end{deluxetable}

\clearpage

\begin{deluxetable}{@{}crcccccc@{}}
\tablecaption{Atmospheric Parameters and Carbon Abundance Estimates
for the Observed Candidates. \label{atmpar}}
\tablehead{Name & V (km/{}s) & $\sigma_{\rm V}$ (km/{}s) & \teff{} (K) & \logg{} (cgs) & \metal{} & \cfe{}\tablenotemark{a} & $\sigma_{\rm \cfe}$}
\tablewidth{0pt}
\tabletypesize{\small}
\startdata

HE~0008$+$0049 & $-$27.6 & 13.3 & 5054 & 4.27 & $-$1.73 & 0.26 & 0.13 \\
HE~0024$-$0550 & 80.6 & 7.2 & 5761 & 4.39 & $-$1.94 & 0.44 & 0.06 \\
HE~0034$-$0011 & $-$173.0 & 18.3 & 6111 & 4.39 & $-$2.16 & 1.73 & 0.13 \\
HE~0035$-$5803 & 78.7 & 25.7 & 6083 & 4.57 & $-$0.65 & 1.00 & 0.05 \\
HE~0053$-$0356 & $-$5.8 & 13.6 & 6004 & 4.39 & $-$1.98 & 1.73 & 0.13 \\
HE~0058$+$0141 & 17.1 & 10.5 & 6670 & 4.57 & $-$0.46 & 0.77 & 0.07 \\
HE~0100$-$4957 & 184.2 & 17.2 & 5050 & 2.61 & $-$2.32 & $-$0.11 & 0.20 \\
HE~0102$-$0004 & $-$106.2 & 13.2 & 6314 & 3.93 & $-$2.20 & 1.19 & 0.19 \\
HE~0118$-$4834 & $-$86.0 & 28.0 & 6015 & 4.50 & $-$2.34 & 2.09 & 0.25 \\
HE~0156$-$5608 & 279.0 & 8.1 & 5431 & 4.32 & $-$2.02 & 0.77 & 0.06 \\
HE~0159$-$5216 & 42.3 & 8.0 & 5413 & 3.80 & $-$1.90 & 0.77 & 0.09 \\
HE~0214$-$0818 & 46.6 & 4.7 & 6379 & 4.34 & $-$1.12 & 1.25 & 0.09 \\
HE~0307$-$5339 & 207.5 & 37.1 & 5689 & 4.32 & $-$1.96 & 1.06 & 0.25 \\
HE~0316$-$2903 & 269.3 & 21.4 & 5528 & 4.43 & $-$2.29 & 1.04 & 0.13 \\
HE~0320$-$1242 & 123.8 & 14.3 & 5745 & 4.39 & $-$0.88 & 0.17 & 0.20 \\
HE~0322$-$3720 & $-$3.5 & 21.3 & 4901 & 4.71 & $-$0.96 & 0.10 & 0.20 \\
HE~0336$-$3948 & 165.3 & 2.7 & 6066 & 4.52 & $-$0.64 & 0.44 & 0.07 \\
HE~0340$-$3933 & $-$1.1 & 6.2 & 6226 & 4.57 & $-$0.28 & 0.06 & 0.20 \\
HE~0345$+$0006 & 17.2 & 13.2 & 5262 & 3.32 & $-$2.50 & 0.18 & 0.06 \\
HE~0405$-$4411 & 126.1 & 11.3 & 6337 & 4.05 & $-$1.14 & 0.89 & 0.13 \\
HE~0414$-$4645 & 92.0 & 24.8 & 6197 & 4.59 & $-$1.02 & 0.77 & 0.06 \\
HE~0440$-$5525 & 87.3 & 25.2 & 6186 & 4.25 & $-$1.18 & 0.52 & 0.19 \\
HE~0444$-$3536 & 181.6 & 16.4 & 5417 & 3.96 & $-$1.57 & 1.16 & 0.13 \\
HE~0449$-$1617 & 116.7 & 15.0 & 5756 & 4.50 & $-$1.07 & 0.06 & 0.20 \\
HE~0451$-$3127 & 342.4 & 26.2 & 5373 & 3.59 & $-$2.97 & 1.13 & 0.13 \\
HE~0500$-$5603 & 156.0 & 23.6 & 4273 & 1.64 & $-$1.61 & $-$0.50 & 0.20 \\
HE~0509$-$1611 & 114.7 & 22.3 & 5279 & 3.80 & $-$1.03 & 0.36 & 0.13 \\
HE~0511$-$3411 & 98.9 & 34.3 & 6055 & 4.57 & $-$0.44 & 0.38 & 0.05 \\
HE~0514$-$5449 & 182.7 & 8.4 & 6414 & 4.16 & $-$0.86 & 0.32 & 0.20 \\
HE~0518$-$3941 & 58.7 & 36.4 & 7153 & 3.34 & $-$0.49 & 1.00 & 0.20 \\
HE~0535$-$4842 & 78.3 & 21.0 & 5910 & 4.43 & $-$0.99 & 0.48 & 0.13 \\
HE~0536$-$5647 & 180.3 & 22.5 & 5417 & 4.02 & $-$1.39 & $-$0.12 & 0.20 \\
HE~0537$-$4849 & 92.9 & 23.3 & 5938 & 4.71 & $-$0.30 & 0.10 & 0.20 \\
HE~0901$-$0003 & 30.4 & 20.0 & 5720 & 4.64 & $-$0.69 & 0.88 & 0.13 \\
HE~0910$-$0126 & 197.4 & 16.9 & 6694 & 3.89 & $-$1.92 & 1.03 & \nodata \\
HE~0912$+$0200 & 83.5 & 15.2 & 5395 & 4.48 & $-$0.75 & $-$0.09 & 0.20 \\
HE~0918$-$0156 & 99.6 & 16.8 & 4237 & 1.61 & $-$1.12 & 0.18 & 0.20 \\
HE~0922$-$0337 & 78.3 & 62.2 & 4943 & 4.14 & $-$1.38 & $-$0.14 & 0.20 \\
HE~0923$-$0323 & 127.9 & 17.7 & 5905 & 4.27 & $-$0.48 & 0.46 & 0.27 \\
HE~0928$+$0003 & $-$133.9 & 70.5 & 4402 & 4.09 & $-$1.14 & $-$0.11 & 0.20 \\
HE~0928$+$0059 & 13.6 & 7.8 & 6598 & 4.39 & $-$0.83 & 0.59 & 0.09 \\
HE~0933$-$0733 & 61.7 & 16.7 & 5982 & 4.39 & $-$1.03 & 0.77 & 0.14 \\
HE~0934$-$1058 & $-$406.9 & 37.1 & 4702 & 3.14 & $-$1.77 & $-$0.50 & 0.25 \\
HE~0948$+$0107 & 514.9 & 5.2 & 5382 & 4.46 & $-$2.14 & 0.20 & 0.06 \\
HE~0948$-$0234 & 147.3 & 54.4 & 6021 & 4.55 & $-$0.44 & 0.38 & 0.19 \\
HE~0950$-$0401 & 144.0 & 14.4 & 6197 & 4.46 & $-$1.62 & 2.01 & 0.13 \\
HE~0950$-$1248 & 87.6 & 18.3 & 5982 & 4.41 & $-$0.30 & 0.10 & 0.20 \\
HE~0951$+$0114 & $-$255.2 & 32.9 & 4882 & 4.61 & $-$1.32 & $-$0.50 & 0.20 \\
HE~1001$-$1621 & $-$10.4 & 26.2 & 5888 & 4.30 & $-$0.92 & 0.17 & 0.20 \\
HE~1002$-$1405 & 99.2 & 9.5 & 6077 & 4.50 & $-$0.44 & 0.38 & 0.05 \\
HE~1007$-$1524 & 96.5 & 18.3 & 6077 & 4.43 & $-$0.64 & 0.46 & 0.19 \\
HE~1009$-$1342 & $-$154.3 & 17.3 & 4242 & 4.30 & $-$1.72 & 0.02 & 0.20 \\
HE~1009$-$1613 & 91.0 & 11.8 & 5883 & 4.34 & $-$0.67 & 0.84 & 0.15 \\
HE~1009$-$1646 & 21.3 & 39.6 & 5883 & 4.46 & $-$0.70 & 0.41 & 0.13 \\
HE~1010$-$1445 & 202.0 & 14.0 & 5117 & 3.43 & $-$0.89 & $-$0.03 & 0.20 \\
HE~1022$-$0730 & 110.3 & 11.4 & 6060 & 4.41 & $-$1.58 & 0.99 & 0.13 \\
HE~1027$-$1217 & 146.6 & 27.5 & 5720 & 2.00 & $-$1.49 & \nodata & \nodata \\
HE~1028$-$1505 & 65.4 & 33.9 & 4886 & 3.84 & $-$0.57 & 0.04 & 0.20 \\
HE~1039$-$1019 & 117.1 & 12.4 & 5878 & 4.55 & $-$0.67 & 0.92 & 0.08 \\
HE~1045$+$0226 & 214.2 & 21.5 & 5109 & 1.90 & $-$3.05 & 2.30 & 0.25 \\
HE~1046$-$1644 & $-$39.5 & 25.6 & 5168 & 3.02 & $-$0.60 & 0.19 & 0.20 \\
HE~1049$-$0922 & $-$34.7 & 30.3 & 5074 & 4.71 & $-$0.63 & 0.33 & 0.20 \\
HE~1049$-$1025 & $-$175.7 & 43.4 & 5634 & 4.50 & $-$0.78 & $-$0.50 & 0.20 \\
HE~1104$-$0238 & 166.8 & 24.0 & 4450 & 1.77 & $-$0.94 & $-$0.5 & 0.05 \\
HE~1110$-$1625 & 123.5 & 17.7 & 5998 & 4.46 & $-$0.30 & 0.11 & 0.20 \\
HE~1112$-$0203 & 8.8 & 41.0 & 4235 & 4.23 & $-$0.91 & $-$0.50 & 0.20 \\
HE~1125$-$1343 & $-$169.4 & 33.2 & 4752 & 4.48 & $-$1.02 & $-$0.03 & 0.20 \\
HE~1129$-$1405 & 189.3 & 41.7 & 5476 & 3.34 & $-$2.02 & 0.60 & 0.09 \\
HE~1132$-$0915 & 51.5 & 41.7 & 5905 & 2.00 & $-$1.56 & \nodata & \nodata \\
HE~1133$-$0802 & 27.5 & 13.9 & 5431 & 4.07 & $-$1.40 & 0.57 & 0.06 \\
HE~1135$-$0800 & 221.3 & 7.8 & 5225 & 3.05 & $-$2.28 & 0.08 & 0.03 \\
HE~1137$-$1259 & 138.1 & 23.5 & 5050 & 4.64 & $-$1.05 & 0.16 & 0.20 \\
HE~1142$-$0637 & 118.4 & 24.3 & 5089 & 4.14 & $-$1.58 & $-$0.04 & 0.20 \\
HE~1146$-$1040 & $-$15.0 & 27.2 & 5382 & 4.57 & $-$1.18 & 0.37 & 0.05 \\
HE~1146$-$1126 & 332.1 & 14.1 & 5062 & 2.84 & $-$2.26 & 0.00 & 0.20 \\
HE~1147$-$1057 & 113.9 & 10.6 & 5971 & 4.52 & $-$0.89 & 0.42 & 0.06 \\
HE~1148$-$1020 & 238.6 & 19.3 & 5835 & 4.39 & $-$1.34 & 0.15 & 0.20 \\
HE~1148$-$1025 & 194.6 & 28.6 & 5792 & 4.52 & $-$0.83 & 0.03 & 0.20 \\
HE~1212$-$1123 & 111.6 & 21.8 & 6503 & 4.07 & $-$1.35 & 0.93 & 0.06 \\
HE~1217$-$1054 & 60.2 & 34.9 & 5156 & 4.57 & $-$0.96 & $-$0.02 & 0.20 \\
HE~1217$-$1633 & 155.6 & 24.6 & 5300 & 3.27 & $-$1.90 & 1.03 & 0.38 \\
HE~1222$-$1631 & 104.6 & 13.7 & 5101 & 3.05 & $-$2.07 & 0.18 & 0.06 \\
HE~1223$-$0930 & 187.0 & 14.0 & 5377 & 3.43 & $-$2.19 & 1.76 & 0.19 \\
HE~1224$-$0723 & 64.9 & 24.7 & 5803 & 4.57 & $-$0.67 & 0.84 & 0.15 \\
HE~1224$-$1043 & 303.0 & 21.9 & 6094 & 3.32 & $-$1.67 & $-$0.08 & 0.20 \\
HE~1228$-$0750 & 353.6 & 14.1 & 6444 & 3.68 & $-$1.60 & 0.40 & 0.20 \\
HE~1228$-$1438 & 176.1 & 19.5 & 4434 & 2.27 & $-$0.92 & $-$0.50 & 0.05 \\
HE~1231$-$3136 & 81.9 & 49.2 & 6279 & 3.59 & $-$1.51 & 0.99 & 0.06 \\
HE~1255$-$2734 & $-$21.2 & 34.6 & 5730 & 4.43 & $-$2.14 & 1.30 & 0.13 \\
HE~1301$+$0014 & 72.0 & 19.5 & 5571 & 3.61 & $-$2.37 & 0.44 & 0.06 \\
HE~1301$-$1405 & 43.0 & 7.9 & 5467 & 3.41 & $-$1.29 & $-$0.11 & 0.20 \\
HE~1302$-$0954 & 145.7 & 22.9 & 5417 & 4.00 & $-$2.30 & 0.77 & 0.13 \\
HE~1311$-$3002 & 213.6 & 27.8 & 5043 & 2.86 & $-$2.39 & 0.64 & 0.09 \\
HE~1320$-$1130 & 220.8 & 41.9 & 6238 & 4.39 & $-$1.62 & 1.37 & 0.08 \\
HE~1320$-$1641 & 84.1 & 14.5 & 4295 & 1.73 & $-$1.03 & $-$0.28 & 0.20 \\
HE~1321$-$1652 & 91.1 & 41.0 & 6169 & 4.61 & $-$1.72 & 2.13 & 0.25 \\
HE~1343$+$0137 & 130.6 & 19.5 & 5808 & 3.23 & $-$1.73 & $-$0.02 & 0.20 \\
HE~1408$-$0444 & 142.5 & 27.9 & 6912 & 3.09 & $-$2.38 & 1.79 & \nodata \\
HE~1409$-$1134 & 92.2 & 12.9 & 6117 & 4.43 & $-$0.37 & 0.23 & 0.20 \\
HE~1410$-$0549 & 103.8 & 25.9 & 6742 & 2.21 & $-$1.70 & \nodata & \nodata \\
HE~1414$-$1644 & 91.8 & 11.8 & 5514 & 4.23 & $-$2.46 & 0.70 & 0.04 \\
HE~1418$-$1634 & 144.3 & 39.1 & 5196 & 2.91 & $-$2.28 & 0.12 & 0.06 \\
HE~1428$-$0851 & 91.3 & 22.4 & 5241 & 1.68 & $-$2.52 & $-$0.48 & \nodata \\
HE~1430$-$1518 & 383.0 & 26.2 & 4289 & 1.93 & $-$1.16 & 0.41 & 0.08 \\
HE~1447$-$1533 & 57.1 & 24.3 & 4234 & 2.07 & $-$1.05 & 0.13 & 0.20 \\
HE~1448$-$1406 & $-$173.8 & 22.7 & 6060 & 2.36 & $-$1.67 & 0.10 & 0.20 \\
HE~1451$-$0659 & $-$22.9 & 57.5 & 4882 & 4.64 & $-$1.47 & $-$0.11 & 0.20 \\
HE~1458$-$0923 & $-$343.0 & 27.3 & 5829 & 4.39 & $-$2.27 & 1.89 & 0.13 \\
HE~1458$-$1022 & $-$61.3 & 18.2 & 5200 & 3.09 & $-$2.22 & 0.47 & 0.06 \\
HE~1458$-$1226 & $-$10.1 & 30.5 & 5504 & 4.68 & $-$1.05 & 0.39 & 0.13 \\
HE~1504$-$1534 & 60.3 & 6.9 & 4246 & 1.41 & $-$0.73 & 0.32 & 0.06 \\
HE~1505$-$0826 & 69.2 & 18.4 & 6730 & 4.39 & $-$0.49 & 0.48 & 0.20 \\
HE~1507$-$1055 & 174.0 & 16.0 & 4267 & 1.66 & $-$1.14 & 0.18 & 0.20 \\
HE~1507$-$1104 & 124.1 & 12.0 & 4492 & 2.91 & $-$0.98 & $-$0.01 & 0.20 \\
HE~1512$+$0149 & $-$65.3 & 41.2 & 4706 & 4.57 & $-$0.94 & $-$0.50 & 0.20 \\
HE~1516$-$0107 & 16.9 & 28.5 & 5720 & 3.77 & $-$2.01 & 0.65 & 0.03 \\
HE~1518$-$0541 & 32.1 & 30.1 & 5217 & 4.68 & $-$1.03 & 0.13 & 0.20 \\
HE~1527$-$0740 & 27.4 & 28.4 & 5679 & 2.00 & $-$1.86 & 0.26 & \nodata \\
HE~1529$-$0838 & 38.7 & 24.6 & 5960 & 4.52 & $-$0.69 & 0.57 & 0.19 \\
HE~2025$-$5221 & 237.5 & 19.7 & 5932 & 4.57 & $-$2.25 & 2.00 & 0.25 \\
HE~2052$-$5610 & 281.7 & 35.8 & 6628 & 4.59 & $-$1.76 & 2.69 & 0.05 \\
HE~2112$-$5236 & 254.5 & 15.0 & 5304 & 3.93 & $-$1.79 & 0.76 & 0.19 \\
HE~2117$-$6018 & $-$205.1 & 30.6 & 5023 & 4.11 & $-$1.93 & $-$0.17 & 0.20 \\
HE~2140$-$4746 & 70.8 & 14.7 & 6111 & 4.36 & $-$1.36 & 0.49 & 0.06 \\
HE~2151$-$0332 & $-$99.6 & 21.0 & 5514 & 4.30 & $-$2.75 & 1.45 & 0.19 \\
HE~2201$-$1108 & $-$115.3 & 19.0 & 6533 & 4.07 & $-$0.95 & 0.79 & 0.19 \\
HE~2207$-$0912 & $-$67.2 & 13.7 & 5824 & 4.30 & $-$2.40 & 0.81 & 0.13 \\
HE~2209$-$1212 & 107.8 & 30.7 & 6432 & 4.43 & $-$0.38 & 0.26 & 0.20 \\
HE~2219$-$1357 & 133.0 & 14.4 & 7082 & 4.25 & $-$0.64 & 0.72 & 0.20 \\
HE~2231$-$0710 & 63.1 & 34.4 & 5704 & 2.86 & $-$0.61 & 0.72 & 0.29 \\
HE~2257$-$5710 & 41.6 & 17.2 & 5343 & 3.89 & $-$2.97 & 1.22 & 0.06 \\
HE~2353$-$5329 & 105.3 & 12.6 & 6509 & 4.57 & $-$1.75 & 2.35 & 0.13 \\

\enddata
\tablenotetext{a}{The \cfe{} values with no errors associated are
upper limits.} 

\end{deluxetable}

\end{document}